\documentclass[10pt,twocolumn,letterpaper]{article}

\usepackage{cvpr}
\usepackage{amsmath,amssymb}
\usepackage{booktabs}
\usepackage{multirow}
\usepackage{array}
\usepackage{graphicx}
\usepackage{url}
\usepackage{xspace}
\usepackage[pagebackref=false,breaklinks=true,colorlinks=true,allcolors=blue]{hyperref}

\def\paperID{0000}
\def\confName{CVPR}
\def\confYear{2026}

\newcommand{\method}{MG-Former\xspace}
\newcommand{\dataset}{CG-Data\xspace}
\newcommand{\R}{\mathbb{R}}
\newcommand{\Lrec}{\mathcal{L}_{\mathrm{rec}}}
\newcommand{\Lalign}{\mathcal{L}_{\mathrm{align}}}
\newcommand{\Lclip}{\mathcal{L}_{\mathrm{CLIP}}}
\title{\method: A Transformer-Based Framework for Music-Driven \\ 3D Conducting Gesture Generation}

\author{Ke Qiu\\
Malou Tech Inc\\
{\tt\small ke.qiu@maloutech.com}
\and
Yawen Qin\thanks{Contribute equally.}\\
South-Central Minzu University\\
{\tt\small qyawen@mail.scuec.edu.cn}
\and
Tianzhi Jia\\
Beijing Jiaotong University\\
{\tt\small jiatianzhi@bjtu.edu.cn}
\and
Xiaole Yang\\
ADVANCE.AI\\
{\tt\small yangxiaole6767@gmail.com}
\and
Kaimin Wang\\
Fudan University\\
{\tt\small kmwang22@m.fudan.edu.cn}
\and
Kaixing Yang\\
Renmin University of China\\
{\tt\small yangkaixing@ruc.edu.cn}
}

\begin{document}
\maketitle

\begin{abstract}
Generating expressive conducting gestures from music is a challenging cross-modal motion synthesis problem: the output must follow long-range musical structure, preserve beat-level synchronization, and remain plausible as a fine-grained 3D human performance. Existing conducting-motion studies are often limited by sparse pose representations, small-scale data, or evaluation protocols that do not directly measure whether music and gesture are mutually aligned. This paper presents \method, a Transformer-based framework for music-driven conducting gesture generation. We introduce a SMPL-parameter data construction pipeline that recovers detailed body motion from conducting videos and forms \dataset, a dataset targeted at professional conducting gestures. Given acoustic descriptors extracted from audio and an initial pose, \method uses a Trans-Temporal Music Encoder and a Trans-Temporal Conducting Gesture Decoder to autoregressively predict SMPL pose parameters. To better assess artistic correspondence, we further build a retrieval-based evaluation model that embeds music and gestures into a shared space and yields FID, modality distance, multi-modality distance, and diversity metrics. Experiments show that \method outperforms dance-generation and conducting-generation baselines, while ablations verify the benefits of the Transformer backbone and the proposed alignment loss.
\end{abstract}

\section{Introduction}

\begin{figure*}[t]
  \centering
  \includegraphics[width=0.86\textwidth]{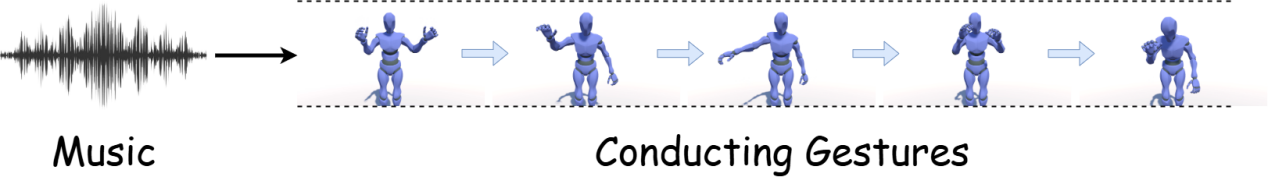}
  \caption{Given a piece of music, \method generates an audio-visually synchronized 3D conducting gesture sequence. The task requires both semantic correspondence to musical phrasing and fine temporal alignment with rhythm and beat changes.}
  \label{fig:teaser}
\end{figure*}

Conducting is a specialized form of embodied musical communication. A conductor does not merely move in time with a score; the gestures encode beat, tempo, dynamics, phrasing, entrance cues, emotional character, and rehearsal intent. Automatic conducting gesture generation therefore has practical value for choir rehearsal, music education, virtual performance, animation, and intelligent tutoring systems. It also offers a compact but demanding benchmark for cross-modal human motion generation: unlike generic gesture synthesis, conducting requires a consistent mapping between musical structure and canonical movement; unlike dance generation, the motion is constrained by professional conventions and should remain interpretable to performers.

Recent music-driven motion generation has progressed rapidly. Music-to-dance systems have moved from recurrent models and GANs~\cite{tang2018dance,qi2019music,sun2021deepdance} to transformer, VQ, and diffusion models capable of long, diverse choreography~\cite{li2021ai,siyao2022bailando,tseng2023edge,li2024lodge}. However, music-to-conducting remains comparatively underexplored. Existing conducting methods have introduced synchronization objectives, diffusion-based generation, and transfer learning under limited data~\cite{liu2022self,zhao2023taming,oh2024transfer}, but the field still faces three persistent limitations. First, the pose representation is often sparse, making it difficult to preserve detailed upper-body rotations, hand trajectories, and body orientation. Second, the music--gesture relationship is not fully modeled at both temporal and semantic levels. Third, common motion metrics do not directly evaluate whether a generated gesture is retrievable from, or discriminative for, the corresponding music.

To address these issues, we propose \method, a Transformer-based framework for music-driven 3D conducting gesture generation. Our design follows a simple principle: music features should first be temporally contextualized, and the gesture decoder should then attend to both previously generated motion and evolving musical context. The Trans-Temporal Music Encoder extracts acoustic dynamics from frame-aligned audio descriptors, while the Trans-Temporal Conducting Gesture Decoder autoregressively predicts SMPL pose parameters~\cite{loper2015smpl,zhou2019continuity}. We train the model with reconstruction and alignment losses so that generated gestures are numerically close to ground truth and semantically close in a learned music--gesture retrieval space.

We also construct \dataset, a dataset pipeline based on video collection, cleaning, pose estimation, anomaly detection, SMPL fitting, and rendering verification. Compared with sparse 3D keypoints, SMPL parameters offer a more detailed and reusable representation for conducting, especially for upper-body rotations and global posture. Finally, we introduce retrieval-based evaluation metrics. Inspired by cross-modal contrastive learning~\cite{radford2021learning}, a retrieval model maps music clips and gesture clips into a shared embedding space; the learned space is then used to compute distributional and pairwise correspondence metrics.

Our contributions are summarized as follows:
\begin{itemize}
  \item We present \dataset, a SMPL-parameter-based data construction pipeline for detailed 3D conducting gesture modeling.
  \item We propose \method, a Transformer-based music-to-conducting framework that improves temporal and semantic correspondence between music and generated gestures.
  \item We design a retrieval-based evaluation protocol that better reflects practical music--gesture matching than purely geometric motion metrics.
  \item We conduct comparison, ablation, and visualization studies demonstrating the effectiveness of the framework, the alignment loss, and the dataset representation.
\end{itemize}

\section{Related Work}

\subsection{3D Dance Generation}
Music-driven dance generation is the closest large-scale neighbor to conducting generation because both tasks map audio to temporally structured 3D motion. Early systems used neural sequence models and adversarial objectives to synthesize beat-aware movement from audio features~\cite{alemi2017groovenet,tang2018dance,qi2019music,sun2021deepdance,huang2021choreography,ren2020self}. Dataset-driven work then improved realism and body detail through large-scale 3D dance corpora and parametric motion representations~\cite{li2021ai,li2023finedance,li2022danceformer}. A second line studies discrete motion memories, transformers, diffusion, and editable generation for longer and more controllable choreography~\cite{siyao2022bailando,siyao2023bailando,tseng2023edge,li2024lodge,li2024lodgepp}. Recent studies further combine music with text or richer control signals, and explore holistic or foundation-model-driven dance synthesis~\cite{gong2023tm2d,li2024multimodal,li2025soulnet,liu2025dgfm}. Group-dance and dance-video methods model dancer interactions, formation coherence, tokenized long motion, refined flow matching, mixture-of-experts generation, and 3D-motion-driven video synthesis~\cite{le2023music,le2023controllable,yang2024codancers,yang2024cohedancers,yang2025megadance,yang2025flowerdance,yang2026tokendance,yang2025macedance}.

These studies provide useful temporal modeling and music-motion alignment tools, but conducting differs from dance in motion semantics and evaluation. Dance emphasizes expressive whole-body choreography, stylistic diversity, and visual novelty; conducting emphasizes beat clarity, cueing, controlled upper-body motion, and readability to musicians. Thus, dance models cannot be directly transferred without handling professional conducting constraints, detailed pose representation, and correspondence-aware evaluation.

\subsection{3D Gesture Generation}
3D gesture generation is another relevant line because it treats human motion as a communicative signal conditioned on speech, text, identity, emotion, or multimodal context. Earlier co-speech methods learn individual gesture styles, trimodal audio-text-speaker context, language-grounded pose forecasting, and semantically aware gesture timing~\cite{ginosar2019learning,yoon2020speech,ahuja2019language2pose,kucherenko2020gesticulator}. Style and dataset-oriented work further shows that gesture synthesis must preserve speaker identity, affect, and social variation rather than only optimize joint error~\cite{alexanderson2020style,liu2022beat,yi2023talkshow,chhatre2024amuse}. More recent diffusion and CLIP-style models improve diversity, controllability, and holistic body-face-hand generation~\cite{zhu2023diffgesture,ao2023gesturediffuclip,yang2023diffusestylegesture,zhang2023diffmotion,chen2024diffsheg}. Masked audio-motion learning, semantic emphasis, and long-horizon correction are also used to reduce mismatch and drift in generated gestures~\cite{zhang2025semtalk,zhang2025echomask,zhang2025globaldiff}. Together, these works show that gesture quality depends on semantic alignment, timing, style, and robustness, not only on local joint accuracy.

Conducting shares this communicative nature but is more constrained than conversational gesture. Co-speech motion can be semantically loose, while conducting must follow meter, onset, phrase, and ensemble cueing conventions. A generated conducting sequence must therefore preserve local synchronization and higher-level musical dynamics. This motivates our retrieval-based evaluation: if a gesture truly matches a music clip, the two should be close in a learned cross-modal embedding space.

\subsection{Conducting Motion Generation}
Conducting motion generation is a young branch of music-driven motion synthesis with a stricter communicative function than dance. Existing work has studied self-supervised music-motion synchronization, diffusion-based conducting generation, fine-grained BiMamba-Transformer diffusion, and transfer learning under limited data~\cite{liu2022self,zhao2023taming,jia2026bitdiff,oh2024transfer}. These methods highlight that conducting requires both local beat alignment and global phrase-level structure.

Despite recent progress, two gaps remain. First, many pipelines still use sparse keypoints or limited pose representations, making detailed upper-body rotations and renderable 3D motion difficult to model. Second, evaluation often focuses on geometric realism or generic distributional distance, while professional usefulness depends on whether a gesture corresponds to the music. \method addresses these issues with a SMPL-based representation and a retrieval-guided metric space for detailed pose and artistic alignment.

\section{Dataset}

\begin{figure*}[t]
  \centering
  \includegraphics[width=0.82\textwidth]{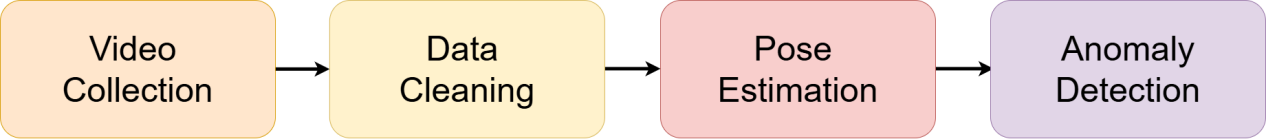}
  \caption{Overview of the \dataset construction pipeline. Videos are collected and cleaned, then processed by pose estimation, anomaly detection, SMPL fitting, and rendering checks to obtain temporally aligned music--gesture pairs.}
  \label{fig:data}
\end{figure*}

\subsection{Motivation}
Existing conducting datasets are limited in scale and representation. Sparse 3D keypoints provide a rough description of body motion, but they discard joint rotations, global orientation, and mesh-level information that are important for downstream rendering and analysis. For conducting, subtle upper-body posture, arm rotation, and wrist trajectory carry musical meaning. We therefore construct \dataset around SMPL pose parameters rather than only keypoint tracks.

\subsection{Data Collection Pipeline}
As shown in Fig.~\ref{fig:data}, the data pipeline contains four major stages. \textbf{Video collection} gathers conducting videos from public performance and rehearsal sources, covering diverse music styles, tempi, and conducting types. \textbf{Data cleaning} removes clips with severe occlusion, unstable framing, non-conducting content, or audio-video desynchronization. \textbf{Pose estimation and SMPL fitting} reconstruct frame-level human pose and maps it to SMPL parameters. We follow robust monocular human reconstruction practices inspired by recent world-grounded recovery methods~\cite{shin2024wham}. \textbf{Anomaly detection} filters frames and clips with implausible body proportions, discontinuous trajectories, failed hand localization, or large temporal jumps.

The final representation stores each gesture frame as a 147-dimensional vector:
\begin{equation}
  g_t = [\tau_t,\theta_t] \in \R^{147},
  \label{eq:pose}
\end{equation}
where $\tau_t \in \R^3$ denotes root translation and $\theta_t \in \R^{144}$ denotes the 6D rotation representation for SMPL joints. The 6D representation is continuous and stable for neural networks~\cite{zhou2019continuity}, avoiding the discontinuities of Euler angles or axis-angle encodings.

\subsection{Audio and Temporal Alignment}
For each clip, audio is resampled and processed with Librosa~\cite{mcfee2015librosa}. We extract a 438-dimensional acoustic descriptor per frame, including MFCCs, MFCC deltas, chroma, tempogram, and onset strength. The audio descriptors and SMPL poses are synchronized at 30 frames per second:
\begin{equation}
  \mathcal{D}=\{(m_{1:T}^{(i)},g_{1:T}^{(i)})\}_{i=1}^{N},
  \quad m_t \in \R^{438},\; g_t \in \R^{147}.
  \label{eq:dataset}
\end{equation}
This alignment is essential: conducting gestures are evaluated not only by whether they look plausible, but also by whether beat, phrase, and cueing information occur at the correct time.

\subsection{Quality Control and Statistics}
After reconstruction, each clip is checked at both the frame level and the sequence level. Frame-level checks remove poses with abnormal joint rotations, large reprojection errors, missing upper-body estimates, or implausible root translation. Sequence-level checks remove clips with long frozen intervals, abrupt global shifts, repeated failed frames, or visible audio-video delay. We also render reconstructed SMPL motion back to video for manual inspection on a subset of clips, because numerical filters may miss errors such as left-right arm swaps or visually unnatural shoulder rotations.

The resulting dataset is designed to cover several axes of variation: choral and solo conducting, slow and fast tempo, lyrical and energetic music, and different camera viewpoints. This diversity helps prevent the model from learning a narrow template of conducting motion. At the same time, clips are normalized into a consistent coordinate convention so that the learning problem focuses on musical and gestural correspondence rather than camera-specific artifacts.

We further keep metadata that is useful for later analysis, including clip duration, coarse musical emotion, conducting type, and reconstruction quality flags. These metadata are not used as direct supervision in the current model, but they make it possible to analyze failure cases by musical condition. For example, fast and passionate clips are more likely to contain large arm movements and motion blur, while slow lyrical clips require smoother trajectories and more stable posture. Keeping such metadata helps separate model limitations from data-quality limitations.

Compared with common dance datasets, conducting data has a different imbalance pattern. Many public conducting videos emphasize frontal upper-body shots, but professional gestures may include subtle torso rotation, preparatory lift, and body-weight transfer that are not obvious from sparse keypoints. This is why we adopt SMPL parameters: the representation makes these details available to the model and also supports consistent rendering for qualitative inspection.

\section{Method}

\begin{figure*}[t]
  \centering
  \includegraphics[width=0.92\textwidth]{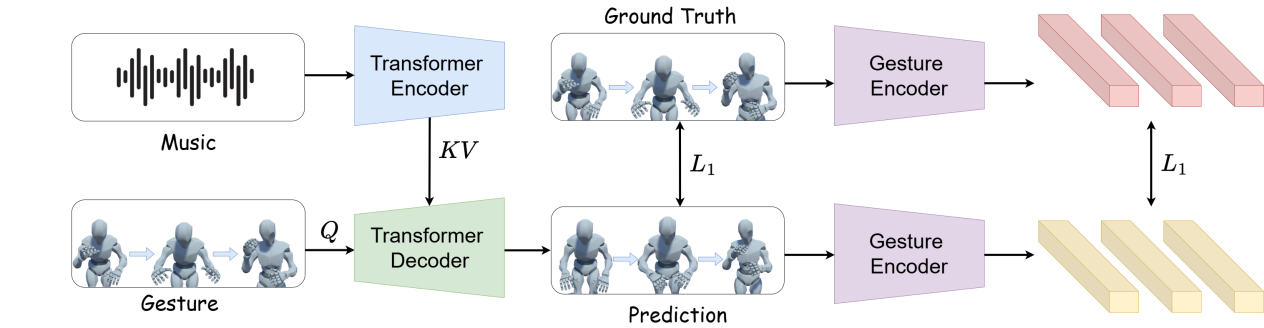}
  \caption{Overview of \method. Music descriptors are encoded by the Trans-Temporal Music Encoder. The gesture decoder autoregressively attends to previous gestures and encoded music features to predict SMPL pose parameters.}
  \label{fig:method}
\end{figure*}

\subsection{Problem Definition}
Given a sequence of music features $m_{1:T}=\{m_t\}_{t=1}^{T}$ and an initial conductor pose $g_0$, the goal is to generate a temporally aligned conducting gesture sequence $\hat{g}_{1:T}=\{\hat{g}_t\}_{t=1}^{T}$. Each $m_t\in\R^{438}$ is an acoustic feature vector and each $\hat{g}_t\in\R^{147}$ is a SMPL pose vector. The generation function is
\begin{equation}
  \hat{g}_{1:T} = F_{\Theta}(m_{1:T}, g_0),
  \label{eq:generation}
\end{equation}
where $\Theta$ denotes learnable parameters. Because conducting depends on both current music and previous motion, we factor $F_{\Theta}$ into a music encoder and an autoregressive gesture decoder:
\begin{align}
  f^m_{1:T} &= E_{\theta}(m_{1:T}), \label{eq:encoder}\\
  \hat{g}_t &= D_{\phi}(f^m_{1:T}, \hat{g}_{<t}, g_0).
  \label{eq:decoder}
\end{align}

\subsection{Trans-Temporal Music Encoder}
The music encoder is built on Transformer self-attention~\cite{vaswani2017attention}. Acoustic descriptors are first projected into a $d$-dimensional hidden space and combined with positional encodings. For an input sequence $X\in\R^{T\times d}$, scaled dot-product attention is
\begin{equation}
  \operatorname{Attn}(Q,K,V)
  = \operatorname{softmax}\!\left(\frac{QK^\top}{\sqrt{d_k}}\right)V,
  \label{eq:attention}
\end{equation}
where $Q=XW_Q$, $K=XW_K$, $V=XW_V$, and $d_k$ is the key dimension. Multi-head attention is then
\begin{equation}
  \begin{aligned}
  \operatorname{MHA}(X)
  &= \operatorname{Concat}(h_1,\ldots,h_H)W_O,\\
  h_i &= \operatorname{Attn}(XW_Q^i,XW_K^i,XW_V^i).
  \end{aligned}
  \label{eq:mha}
\end{equation}
By attending to different temporal positions simultaneously, the encoder captures rhythm, onset, harmonic change, and emotional progression over long windows. This is important for conducting because gestures often anticipate musical events rather than merely react to the current audio frame.

We use temporal windows rather than isolated frames because conducting gestures are highly anticipatory. For example, a preparatory motion may begin before a musical entrance and resolve at the beat. A frame-wise regressor can fit local pose statistics but may fail to model this preparation--arrival relationship. The Transformer encoder gives every music token access to the surrounding phrase context, allowing the decoder to condition a current gesture on both local rhythmic events and longer musical structure.

\subsection{Trans-Temporal Conducting Gesture Decoder}
The decoder predicts SMPL pose parameters autoregressively. At time $t$, it uses two attention pathways. The first pathway performs causal self-attention over previously generated gestures:
\begin{equation}
  z_t^{g} = \operatorname{MHA}_{\mathrm{self}}(\hat{g}_{<t}),
  \label{eq:selfattn}
\end{equation}
which encourages motion continuity and coherent gesture phrasing. The second pathway performs cross-attention from the gesture state to encoded music features:
\begin{equation}
  \begin{aligned}
  z_t^{m} &=
  \operatorname{Attn}(Q_t^{g},K^m,V^m),\\
  Q_t^{g}&=z_t^{g}W_Q,\quad
  K^m=f^m_{1:T}W_K,\quad
  V^m=f^m_{1:T}W_V.
  \end{aligned}
  \label{eq:crossattn}
\end{equation}
The final pose prediction is produced by a feed-forward head:
\begin{equation}
  \hat{g}_t = \operatorname{MLP}([z_t^{g};z_t^{m}]).
  \label{eq:head}
\end{equation}
The self-attention branch keeps the generated movement physically and stylistically consistent, while the cross-attention branch aligns the movement with evolving musical context.

\subsection{Training Objectives}
We use a reconstruction loss and an alignment loss. The reconstruction loss penalizes frame-level deviation from ground-truth SMPL parameters:
\begin{equation}
  \Lrec =
  \frac{1}{T}\sum_{t=1}^{T}
  \|\hat{g}_t-g_t\|_1.
  \label{eq:rec}
\end{equation}
This objective encourages accurate pose, stable rotations, and smooth local trajectories.

The alignment loss compares generated and ground-truth gestures in the embedding space of the retrieval model described in Sec.~\ref{sec:evaluation}. Let $R_g(\cdot)$ be the gesture encoder of the retrieval model. We define
\begin{equation}
  \Lalign =
  1-
  \frac{
  \langle R_g(\hat{g}_{1:T}), R_g(g_{1:T}) \rangle
  }{
  \|R_g(\hat{g}_{1:T})\|_2 \|R_g(g_{1:T})\|_2
  }.
  \label{eq:align}
\end{equation}
The overall objective is
\begin{equation}
  \mathcal{L}=\Lrec+\lambda\Lalign,
  \label{eq:total}
\end{equation}
where $\lambda$ balances geometric fidelity and semantic correspondence. The alignment term improves music--gesture consistency by using a learned space where gestures with similar musical meaning are close.

The two losses play complementary roles. The reconstruction loss preserves exact pose supervision and stabilizes training, especially for common beat patterns. The alignment loss is more tolerant of stylistic variation: two gestures may not be identical frame by frame, yet both can express the same musical phrase. This property is useful in conducting, where different conductors may choose slightly different trajectories while still communicating the same cue.

In practice, we apply the alignment loss after the retrieval encoders have learned a stable shared embedding. This avoids forcing the generator to follow an unstable metric space early in training. The retrieval feature acts as a high-level perceptual proxy: it does not replace frame-level supervision, but it encourages generated motion to preserve the musical identity of a clip. This is particularly helpful for phrase-level conducting gestures, where small local pose differences may be acceptable but mismatched musical intent is not.

The autoregressive design also benefits from the loss combination. Pure reconstruction can make the decoder overfit to average transitions and produce conservative motion, while pure contrastive alignment would be too weak to supervise precise joint rotations. By combining them, the model learns both the physical trajectory of conducting motion and the semantic correspondence between music and gesture. The balance is important because professional conducting requires restrained motion in some passages and high-amplitude cueing in others.

\subsection{Inference and Implementation}
At inference time, \method begins with $g_0$ and generates poses one frame at a time. The previously predicted pose is fed back into the decoder, and the music encoder output remains available for the entire clip. We train on 10-second clips for 500 epochs using AdamW with learning rate $4\times10^{-3}$ and a cosine schedule. Both encoder and decoder contain 6 Transformer layers, 8 attention heads, hidden dimension 512, and dropout 0.1.

\section{Retrieval-Based Evaluation}
\label{sec:evaluation}

\begin{figure}[t]
  \centering
  \includegraphics[width=\columnwidth]{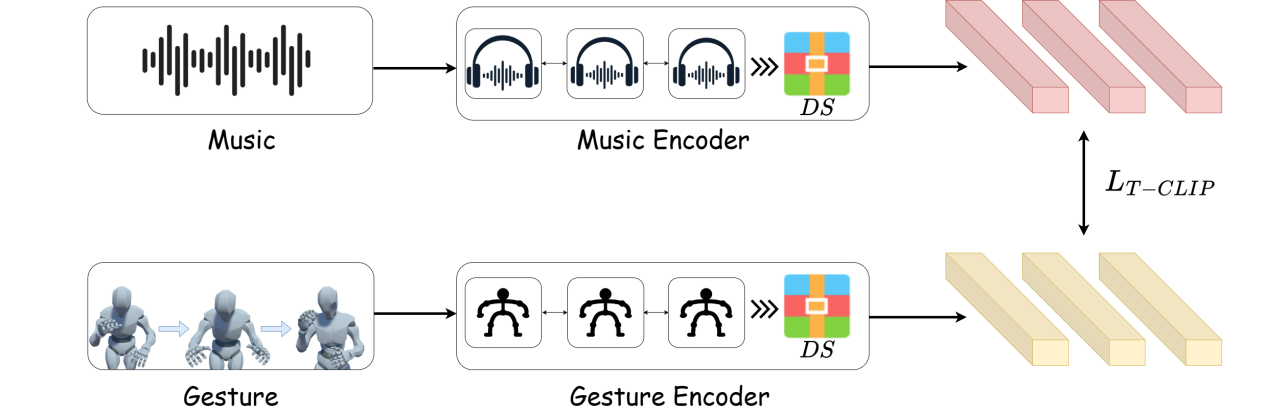}
  \caption{Retrieval model used for evaluation. A music encoder and a gesture encoder map paired clips into a shared embedding space. The resulting space supports music-to-gesture retrieval, gesture-to-music retrieval, and correspondence-aware metrics.}
  \label{fig:retrieval}
\end{figure}

\subsection{Motivation}
Common motion-generation metrics evaluate realism, diversity, or distance to ground truth, but conducting requires more: the gesture should match the music. We therefore formulate evaluation as a bidirectional retrieval problem. If generated gestures correctly express the corresponding music, music embeddings and gesture embeddings should be close in the learned space.

\subsection{Retrieval Model}
The retrieval model contains a Music Encoder $R_m$ and a Gesture Encoder $R_g$. The music branch uses the same 438-dimensional acoustic descriptors as \method, followed by temporal processing and downsampling blocks. The gesture branch uses SMPL 6D pose parameters. Both branches output normalized embeddings:
\begin{equation}
  u_i = \frac{R_m(m_i)}{\|R_m(m_i)\|_2},
  \quad
  v_i = \frac{R_g(g_i)}{\|R_g(g_i)\|_2}.
  \label{eq:embeddings}
\end{equation}
For a batch of $B$ paired clips, the similarity matrix is
\begin{equation}
  s_{ij}=\frac{u_i^\top v_j}{\tau},
  \label{eq:similarity}
\end{equation}
where $\tau$ is a temperature parameter. We train the model with symmetric CLIP-style contrastive loss~\cite{radford2021learning}:
\begin{align}
  \mathcal{L}_{m\rightarrow g}
  &= -\frac{1}{B}\sum_{i=1}^{B}
  \log \frac{\exp(s_{ii})}{\sum_{j=1}^{B}\exp(s_{ij})},\\
  \mathcal{L}_{g\rightarrow m}
  &= -\frac{1}{B}\sum_{i=1}^{B}
  \log \frac{\exp(s_{ii})}{\sum_{j=1}^{B}\exp(s_{ji})},\\
  \Lclip &= \frac{1}{2}
  \left(\mathcal{L}_{m\rightarrow g}+\mathcal{L}_{g\rightarrow m}\right).
  \label{eq:clip}
\end{align}

\subsection{Metrics}
After training, we extract embeddings for real gestures, generated gestures, and music clips. We report four metrics:
\begin{itemize}
  \item \textbf{FID}: Fréchet distance between the distribution of generated gesture embeddings and real gesture embeddings.
  \item \textbf{M-Dist}: average Euclidean distance between generated gesture embeddings and ground-truth gesture embeddings.
  \item \textbf{MM-Dist}: average Euclidean distance between generated gesture embeddings and the corresponding music embeddings.
  \item \textbf{Div}: average pairwise distance among generated gesture embeddings, reflecting diversity.
\end{itemize}
The retrieval model is trained for 50 epochs with Adam, learning rate $4\times10^{-4}$, $\beta=[0.5,0.99]$, 6 temporal blocks, 8 attention heads, hidden dimension 256, and dropout 0.1. We split \dataset into training and testing sets using a 7:3 ratio.

\subsection{Why Retrieval Metrics Fit Conducting}
The retrieval formulation gives a practical approximation of a question that musicians naturally ask: does this gesture belong to this music? A generated sequence may have smooth body motion and reasonable diversity, but if it indicates a strong entrance during a quiet transition, it is artistically mismatched. Conversely, a gesture may differ from the exact ground-truth trajectory while still conveying the correct meter and phrase. Retrieval metrics capture this middle ground by learning correspondence from real paired clips instead of requiring one deterministic target gesture.

\section{Experiments}

\subsection{Settings}
We evaluate \method on \dataset. Each model is trained and evaluated on temporally aligned 10-second music--gesture clips. We compare against FACT~\cite{li2021ai}, a representative music-to-dance baseline, and VirtualConductor, a conducting-generation baseline. Since dance and conducting have different motion distributions, the comparison is intended to test whether the proposed conducting-specific design better matches the professional gesture domain.

All methods are evaluated with the same retrieval model and the same test split. For fair comparison, generated clips are converted into the same SMPL-based representation before metric computation. We report the mean performance over the test set and use identical audio features for all models. This protocol isolates the impact of the generative model rather than differences in preprocessing or evaluation features.

\begin{table}[t]
  \centering
  \caption{Quantitative comparison on \dataset. Lower is better for FID, M-Dist, and MM-Dist; higher is better for Div.}
  \label{tab:comparison}
  \resizebox{\columnwidth}{!}{
  \begin{tabular}{lcccc}
    \toprule
    Method & FID $\downarrow$ & M-Dist $\downarrow$ & MM-Dist $\downarrow$ & Div $\uparrow$\\
    \midrule
    Ground Truth & 0.00 & 0.00 & 21.53 & 21.65\\
    FACT~\cite{li2021ai} & 115.76 & 19.66 & 22.29 & 20.77\\
    VirtualConductor & 100.81 & 19.24 & 22.21 & 20.64\\
    \method & \textbf{91.73} & \textbf{18.47} & \textbf{22.18} & \textbf{20.87}\\
    \bottomrule
  \end{tabular}}
\end{table}

\subsection{Comparison Study}
Table~\ref{tab:comparison} shows that \method achieves the best FID, M-Dist, MM-Dist, and Div among compared generative methods. The improvement in FID indicates that the generated gesture distribution is closer to real conducting motion. The lower M-Dist shows better correspondence to ground-truth gestures, while the lower MM-Dist indicates stronger music--gesture alignment in the retrieval embedding space. The diversity score remains competitive, suggesting that the model does not collapse to a single conservative motion pattern.

\begin{figure*}[t!]
  \centering
  \includegraphics[width=0.90\textwidth]{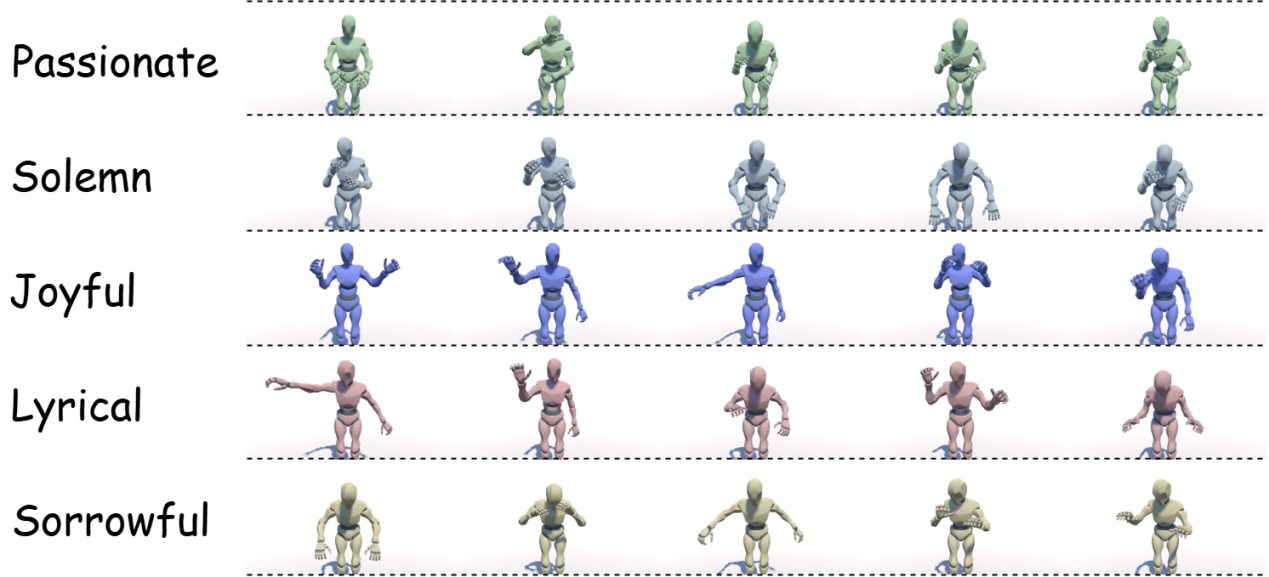}
  \caption{Generated conducting gestures across diverse musical emotions, including passionate, solemn, joyful, lyrical, and sorrowful clips.}
  \label{fig:emotion}
\end{figure*}

\begin{figure*}[t!]
  \centering
  \includegraphics[width=0.84\textwidth]{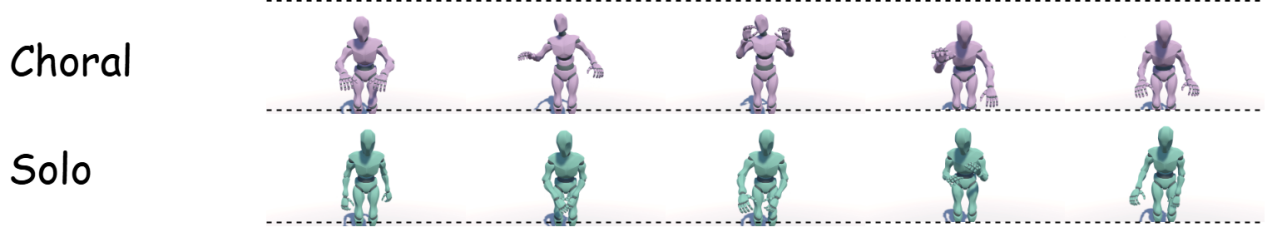}
  \caption{Generated gestures for choral and solo conducting scenarios. The visualizations show that \method can adapt gesture range and posture to different conducting contexts.}
  \label{fig:type}
\end{figure*}

\begin{table}[t]
  \centering
  \caption{Ablation on model backbone.}
  \label{tab:backbone}
  \resizebox{\columnwidth}{!}{
  \begin{tabular}{lcccc}
    \toprule
    Backbone & FID $\downarrow$ & M-Dist $\downarrow$ & MM-Dist $\downarrow$ & Div $\uparrow$\\
    \midrule
    Ground Truth & 0.00 & 0.00 & 21.53 & 21.65\\
    RNN & 137.61 & 19.57 & 22.30 & 19.97\\
    LSTM & 123.07 & 19.64 & 22.28 & \textbf{20.94}\\
    Transformer & \textbf{91.73} & \textbf{18.47} & \textbf{22.18} & 20.87\\
    \bottomrule
  \end{tabular}}
\end{table}

\begin{table}[t]
  \centering
  \caption{Ablation on alignment loss (AL).}
  \label{tab:loss}
  \resizebox{\columnwidth}{!}{
  \begin{tabular}{lcccc}
    \toprule
    Method & FID $\downarrow$ & M-Dist $\downarrow$ & MM-Dist $\downarrow$ & Div $\uparrow$\\
    \midrule
    Ground Truth & 0.00 & 0.00 & 21.53 & 21.65\\
    w/o AL & 99.82 & 18.78 & \textbf{22.17} & \textbf{20.91}\\
    \method & \textbf{91.73} & \textbf{18.47} & 22.18 & 20.87\\
    \bottomrule
  \end{tabular}}
\end{table}

\subsection{Ablation Study}
Table~\ref{tab:backbone} compares RNN, LSTM, and Transformer backbones. The Transformer obtains a large FID improvement, which supports the need for long-range temporal modeling. Conducting gestures often prepare for upcoming musical events; self-attention can connect distant rhythmic and phrase cues more directly than recurrent models.

Table~\ref{tab:loss} evaluates the alignment loss. Removing the alignment loss degrades FID and M-Dist, indicating that retrieval-space supervision helps the generator produce gestures closer to real and paired conducting motion. MM-Dist changes only slightly, suggesting that the retrieval metric is sensitive to both the learned embedding space and the generator's frame-level accuracy. Overall, the alignment loss improves the balance between realism and correspondence.

\subsection{Error Analysis}
We observe three common failure modes. First, the model may underestimate very large gestures in highly energetic music, producing plausible but slightly conservative arm amplitudes. Second, fast transitions can cause short temporal lag when the music contains sudden changes that are rare in the training set. Third, reconstruction noise in the source SMPL data can occasionally propagate into generated wrist or shoulder motion. These failures suggest that future work should combine stronger hand-aware reconstruction, longer musical context, and data balancing across tempo and emotion categories.

A closer look at the ablations suggests that the Transformer backbone mainly improves distributional realism and semantic matching, while the alignment loss mainly regularizes the generated sequence toward musically meaningful gestures. This distinction is useful for future model design. If the goal is long-form stability, stronger temporal architectures or diffusion-style refinement may be more beneficial. If the goal is artistic correspondence, retrieval or contrastive objectives should be strengthened, possibly with phrase-level labels or conductor annotations.

The visualization results also show that qualitative evaluation should not be reduced to a single screenshot. Conducting is inherently temporal: a static frame can show pose plausibility, but it cannot reveal whether the preparation, ictus, rebound, and recovery occur at the right musical moments. For this reason, our retrieval-based metrics are best interpreted together with rendered motion clips and expert inspection. The current paper uses frame-strip visualizations for compact presentation, but the underlying representation supports full video rendering.

\subsection{Visualization Study}
Figures~\ref{fig:emotion} and~\ref{fig:type} provide qualitative examples. Across emotional categories, \method generates different gesture amplitudes and postural tendencies: passionate music tends to produce larger arm trajectories; solemn and lyrical clips lead to more restrained and continuous motion; joyful clips show more open movements. Across conducting types, the model adapts to choral and solo contexts while preserving the temporal structure of the input music. These observations are consistent with the quantitative retrieval metrics.

\section{Conclusion}
We presented \method, a Transformer-based framework for music-driven 3D conducting gesture generation. The method uses temporally contextualized music features and an autoregressive gesture decoder to predict SMPL pose parameters. We also introduced a SMPL-based dataset construction pipeline and a retrieval-based evaluation protocol for measuring music--gesture correspondence. Experiments on \dataset show that \method outperforms existing baselines and benefits from both the Transformer backbone and alignment loss.

The results suggest that detailed pose representation, long-range temporal attention, and correspondence-aware evaluation are all important for conducting generation. Still, the current system inherits limitations from monocular reconstruction and does not explicitly model baton motion, finger articulation, or very long musical structures. Future work should explore hand-aware reconstruction, long-form generation, and musician-in-the-loop assessment. We hope this work encourages further study of conducting as a fine-grained, musically grounded human motion generation task.

{\small
\bibliographystyle{ieeenat_fullname}
\bibliography{references}
}

\end{document}